\documentclass[singlecolumn,showpacs,preprintnumbers,amsmath,amssymb,superscriptaddress]{revtex4}
\usepackage{graphicx}
\usepackage{dcolumn}
\usepackage{bm}
\usepackage{longtable}

\newcommand{\bea}{\begin{eqnarray}}
\newcommand{\eea}{\end{eqnarray}}

\usepackage{amsfonts}
\usepackage{amssymb}
\usepackage{amsmath}
\usepackage{amsthm}
\usepackage{graphicx}

\newcommand{\be}{\begin{equation}}
\newcommand{\ee}{\end{equation}}

\def\Be'{\beta_\mu^{'}}

\def\<{\bigl\langle}
\def\>{\bigr\rangle}

\begin{document}


\title{The ''Life Potential'': a new complex algorithm to assess ''Heart Rate Variability'' from Holter records for cognitive and diagnostic aims.
\newline
\newline
\small{Preliminary experimental results showing its dependence on age,  gender and health conditions.}}

\author{\textbf{Orazio Antonio Barra}}
\affiliation{Department of Environmental and Chemical Engineering, DIATIC, University of Calabria (UNICAL), Via P. Bucci, Rende, Cosenza (Italy)}
\affiliation{International Polytechnic ''Scientia et Ars'' (POLISA) - Palazzo Accademie, Largo Intendenza 1, Vibo Valentia (Italy)}
\affiliation{Email:  orazioantonio.barra@unical.it (Corresponding Author)}

\author{\textbf{Luciano Moretti}}
\affiliation{Department of Cardiology and Hemodynamics, Clinic and Cardiology Research Unit (URCC),  Giulio Bazzoni Hospital - Ascoli Piceno   (Italy) }
\affiliation{Email: luciano.moretti1@sanita.marche.it}

\date{\today}

\pacs{87.57R-, 05.45.Pq, \ 87.19ug, \ 05.10.-a}

%
%
%
%
%
%
%
%
%
%
%
%
%
%
%
%
%
%

\begin{abstract}
Although HRV ( Heart Rate Variability) analyses have been carried out for several decades, several limiting factors still make these analyses useless from a clinical point of view. The present paper aims at overcoming some of these limits by introducing the "Life Potential" (BMP), a new mathematical algorithm which seems to exhibit surprising cognitive and predictive capabilities. BMP is defined  as a linear combination of five HRV Non-Linear Variables, in turn  derived from the thermodynamic formalism of chaotic dynamic systems. The paper presents experimental measurements of BMP (Average Values and Standard Deviations) derived from 1048 Holter tests, matched in age and gender , including a control group of 356 healthy subjects. The main results are: (a) BMP always decreases when the age increases, and its dependence on  age and gender  is well established; (b) the shape of the age dependence within "healthy people" is different from that found in the general group: this behavior provides evidence of possible illness, and seems to justify the attribute of "life potential".  A simplified  "standard" procedure is then supplied to compute  BMP values for a given subject, whereas  BMP dependence on clinical status is now under investigation to make BMP  a useful clinical tool.
\newline
{\em \textbf{Keywords}:} Heart Rate Variability; HRV; Holter RR Data; Non Linear Variables; Chaotic Dynamic Systems; Life Potential; Approximate Entropy; Sample Entropy; Detrended Fluctuations Analyses; Correlation Dimension; Gender Dependence; Age Dependence; Diagnostic Value; Experimental Measurements; ANOVA.
\end{abstract}

\maketitle

\section{Introduction}

Since the first paper of Akselrod et al. \cite{1} published on Science on $1981$, HRV (Heart Rate Variability) studies have shown a considerable  potential to assess the role of autonomic nervous system fluctuations in normal healthy individuals as well as in patients with various cardiovascular and non-cardiovascular disorders. HRV studies should enhance our understanding of physiological phenomena, actions of medical drugs, and disease mechanisms. But, apart a few cases, such as the evaluation of risks of sudden death after a Myocardial Infarction, long term research efforts   are still needed to determine the sensitivity, specificity, and predictive value of HRV in identifying  individuals at risk for subsequent morbidity and mortal events.  Nowadays, the five main factors limiting the development of  HRV-based diagnostic methodologies are:
\begin{itemize}
\item	linear analyses are more diffused than the more appropriate non-linear approaches
\item	HRV techniques are used in practice to analyze heart diseases only, rather than a more complex body's equilibrium
\item	statistical analyses of  HRV variables are normally carried out by comparing Holter data  recorded once-in-a-lifetime on  different patients but as yet no rigorous  analysis of the HRV evolution with time for a given patient is  available
\item	no comparative analysis or "cross analysis" between HRV measurements and other simultaneous acoustic, electromagnetic or "imaging" measurements has yet been carried out to support or  to refuse HRV results
\item	despite of some international activities on-going since $1996$\cite{2}, no systematic approach has yet been  developed, no  measurements and classification procedure has been standardized  and no Common Reference Data-Base (CRDB) is yet available to the scientific community, in order  to validate new diagnostics tools.
\end{itemize}

Within the above framework, a comprehensive research  activity "MATCH" ( Mathematical Advanced Tools to Catch  Heart Rate Variability ) has been carried out  since $2012$  at the Italian UNICAL and POLISA jointly with the Ascoli Piceno Hospital-URCC  to overcome the aforesaid main limits. The main tasks of MATCH are:
\begin{enumerate}
\item	introducing {\em "\textbf{non-linear variables analyses}"}, which might eventually generate new overall predictive-diagnostic algorithms;
\item	introducing a {\em "\textbf{systematic approach}"} to link HRV analysis with a more general clinical characterization of patients;
\item	introducing HRV {\em "\textbf{time-evolution analyses}"} on the same patients: these analyses, on mid  (months) and  long (years) time scales, could confirm or reject several of the conclusions reached with once-in-a-lifetime  analyses, and could also help assessing the effects of pharmaceutical drugs or setting-up new diagnostic tools, as well as suggest warning procedures;
\item	introducing  {\em "\textbf{cross analyses}"} between  two or more different independent methodologies, i.e. among HRV, BRSA (Baro-Reflex Sensitivity Analysis), PDSP (Phonocardiography Digital Signal Processing) and RIAC (Radio-Isotopic Angio-Cardiography);
\item	promoting the birth of CRDB, with certified formats and procedures.
\end{enumerate}

The present paper is focused on the first item above, whereas items $2,3,4,5$ will be dealt with by other short and mid-term forthcoming communications. To this purpose, a new function - hereinafter referred as the "Life Potential" \cite{3}  (BMP, from the initials of the authors)  - has been defined as a {\em linear combination of five HRV Non-Linear Variables: the combination coefficients are free parameters which can be used  "to tune" the function towards different objectives}. The simplest case , where the $5$ coefficients are equal to $1$, is called \textbf{"standard case"}.

\bigskip

The objectives  of the present work are:
\begin{itemize}
\item	To provide  BMP reference values for healthy people, in the "standard case"
\item	To show the capability of BMP to discriminate  between healthy and non-healthy people
\item	To provide  the BMP {\em "alfa"}  and {\em "omega"} values i.e. the BMP upper and lower limits for  young-healthy people and over 80 ill people respectively
\item	To show and to analyze the dependence of BMP on the patient age and gender
\end{itemize}

\textbf{As far as "Ethics" is concerned}, although the work can be considered an "Observational Retrospective Study" (without any risk of harm for patients) carried out on  historical files  available in the Ascoli Piceno Hospital [of "heart failure clinic" and "health fitness certification" for sport practice purposes],  it complies with the principles laid down in the \textbf{"Declaration of Helsinki"}  , and the whole research was conducted with the understanding and the consent of the human subjects and \textbf{with the specific approval  of the Ethical Committee of Regione Marche}.

\section{Materials and Methods}

The ''Life Potential'' for a subject $P$, $BMP_P$, is defined as:
\be
BMP_P = \frac{[ C_{1P} - Ap En_P + C_{2P} \cdot SampEn_P + C_{3P} \cdot (DFA\alpha_1)_P + C_{4P} \cdot (DFA\alpha_2)_P + C_{5P}\cdot D2_P ]}{BMP_H^{AV}}-SF_P
\ee
where
\begin{itemize}

\item $BMP_H^{AV} =$  Reference Potential Value for ''young healthy people'', as defined in the following Sec.$III.A$;

\item $C_{1P}...C_{5P} =$  Set of calibration coefficients, sometimes referred to as ''tuner'' vector;

\item $SF_P =$  Scale (or ''normalization'') factor useful to avoid BMP values $>1$.

\end{itemize}
are free parameters, and $AnEn_P$, $SampEn_P$, $(DFA_{\alpha1})_P$, $(DFA_{\alpha2})_P$, and $D2_P$ represent a set of non-linear variables more and more employed in the HRV investigations. They respectively are: the Approximate Entropy, the Sample Entropy, the Detrended Fluctuations Analysis (DFA): Short Term Fluctuation Slope; the DFA: Long Term Fluctuation Slope; the Correlation Dimension of the RR Data distribution. All the variables above are defined according to the most diffuse standards in literature \cite{4,5,6,7,8}.
\newline
The correct nomenclature for $BMP_P$ is
$$
BMP_P[C_{1P}, C_{2P}, C_{3P}, C_{4P}, C_{5P}, SF_P, BMP_H^{AV}]
$$
The case with $C_{1P}\equiv C_{2P}\equiv C_{3P}\equiv C_{4P}\equiv C_{5P} \equiv 1$ and $SF_P=0.1$ is named the ''\textbf{standard case}'', thus:
\be
BMP_P [1,1,1,1,1,0.1, BMP_H^{AV}] = {[ApEn_P +  SampEn_P + (DFA_{\alpha1})_P + (DFA_{\alpha2})_P +  D2_P] / (BMP_H^{AV})} - 0.1
\ee
BMP is a dimensionless parameter - being the ratio between two functions with the same dimensions - and it ranges, in the standard case,  between 0 and 1;  it  represents \textbf{the ratio between the  life potential of a patient and the life potential of an average healthy young person}.

\bigskip

The results shown in this paper have been obtained by  elaborating data from 1148 Holter Tests ($54$ per cent of Males) - 356 on "Healthy" subjects, and 792 on "Non-Healthy" subjects, as defined hereinafter - carried out at the Ascoli Piceno Hospital with  MORTARA 12-channels devices, which provided some 100 million raw RR Data,  with an average of 5 per cent - and never greater than 10 per cent - of data rejected because of artifacts and mis-triggers.  The duration of all recordings was 24 hours. The raw RR data have been handled by making use of two Computer Codes developed by Vibo V. Polytechnic (POLISA): the first one, based on MATHLAB algorithms, is employed to evaluate the HRV complete set of linear time-domain and frequency domain variables as well as all the HRV non-linear variables for each patient \cite{9 - 23} ; the second one is employed to compute the $BMP_P$  values for any set  of  $C_{1P}, C_{2P}, C_{3P}, C_{4P}, C_{5P}$  coefficients, taking into account in each case the most appropriate $BMP_H$ value.

\bigskip

The patients examined represent a typical sample of the population undergoing a cardiac Holter test in an hospital, so that  all sorts of human, social and clinical characteristics are included; they have been  divided by  age, gender, and clinical status. Among them two special groups have been considered ,  respectively \textbf{"healthy people" (HP)}  and \textbf{"young and healthy people" (YHP)} .  The \textbf{HP group} is made up of people satisfying the following two constraints:
\begin{itemize}
\item	Not  suffering  any apparent serious disease
\item	Not undergoing any important therapy
\end{itemize}
The \textbf{YHP group} is made up by people who satisfy the two constraints above,  have a regular engagement in  sports or other similar physical activity  and  are younger than 25 years: this   group is used to calculate the  aforesaid Reference \textbf{Potential Value, $BMP_H^{AV}$ introduced in equations $(1)$ and $(2)$}, and discussed in Sec. $III.A$.

\bigskip

All the results presented  \textbf{in this  paper are calculated in the ''standard case''}  defined in eqn. $(2)$.
\textbf{Results regarding healthy people, and/or those concerning  differences in age and gender, are presented in details in Sec.$III$}. Other results,  aimed at discriminating between different clinical diseases by means of Ci patterns different from the standard case,  will be the object of future work,  once a large enough Data Base  is available and adequate  protocols and data collection procedures are standardized.

\section{Results and Discussion}

\subsection{The Reference Values $BMP_H$}

\textbf{The first set of activities dealt with the YHP group}: this was a necessary step to estimate the Reference \textbf{Potential Values} $BMP_H^{AV}$  which must be considered in $(2)$ to evaluate the BMP value for a given subject.

\bigskip

The $BMP_H^{AV}$  reference value is  defined as the \textbf{average value of the $BMP_{H^i}$ values} ( " i" ranging between 1 and "n", "n" being the number of people forming the YHP group), where each  $BMP_{H^i}$value refers to the i-th people forming part of the YHP group, and it is defined as :
\be
BMP_{H^i} = [C_{1P} \cdot  ApEn_{H^i} + C_{2P}\cdot  SampEn_{H^i} + C_{3P} \cdot  (DFA\alpha_1)_{H^i}
+ C_{4P} \cdot (DFA\alpha_2)_{H^i} + C_{5P} \cdot D2_{P^i}]      		
\ee
Therefore ,  in the "standard case",  it is  simply given by:
\be
BMP_{H^i} = [ApEn_{H^i} + SampEn_{H^i} + (DFA\alpha1)_{H^i} + (DFA\alpha_2)_{H^i} + D2_{P^i}]
\ee
and  \textbf{it was measured} from the five non-linear HRV variables calculated from  Holter Raw RR Data - both average values and standard deviation values -  for the YHP population, subdivided in three groups according to gender: \textbf{both sexes, men only and women only}.

The information regarding \textbf{"both sexes"} must be privileged for \textbf{general scientific studies} and considerations, whereas, for clinical practice, the most detailed data regarding "men" or "women" seem to be more reliable.

In non-standard cases  the  $BMP_H^{AV}$ value, as well as  its standard deviation amplitude,   vary because of the different weights given to the various HRV variables: therefore, before BMP is evaluated, its reference value $BMP_H^{AV}$ must be recalculated , restarting from the five non-linear HRV variables measured for "YHP" group but considering the proper $C_{1P}...C_{5P}$    set of calibration coefficients characterizing  the selected non standard case.

Furthermore, the results  introduced in the present paper  have been evaluated from the most complete data base of  experimental measurements available today but they should be periodically updated in parallel with the growth of this data base.

\bigskip

Because of the problems above, \textbf{$BMP_H^{AV}$ values for standard case and for some of the most significant non-standard cases will be introduced and discussed in details in a forthcoming paper}.

\bigskip

Once the experimental $BMP_H^{AV}$ values have been evaluated for the aforesaid three groups of YHP population (both sexes, men only and women only),  the dependences of BMP  on age and gender ,  \textbf{both for "all people" and HP , have  been studied , and the relevant results are introduced and discussed in the following sec.$III.B$.}.

\subsection{BMP values and their dependence on age and gender}

Then, after the evaluation of the $BMP_H^{AV}$ values, the available data-base of  Holter RR data has been used to evaluate \textbf{the dependences of the life potential BMP on the people age and gender}. Patients have thus been divided in \textbf{8 age groups}:
\begin{center}
Group 1 :   $<$ 21 years old, \ \ Group 2 :  21-30 years old, \ \ Group 3 :  31-40 years old, \ \ Group 4 :  41-50 years old

\bigskip

Group 5 :  51-60 years old, \ \		 Group 6 :  61-70 years old, \ \ Group 7 :  71-80 years old,  \ \ 		Group 8 :   $>$ 80 years old.
\end{center}
\textbf{In a first phase,  only healthy people (HP)  were considered} (this case hereinafter  is referred as \textbf{"Healthy"}).  \textbf{In a second phase}, no distinction was made between  "healthy" and "non-healthy" people and \textbf{the whole experimental population was considered} (this case hereinafter  is referred as \textbf{"General"}).

\bigskip

The whole set of results is shown in \textbf{Figures 1,2,3} which summarize the results regarding the \textbf{BMP  average values and the relevant standard deviation values $(\sigma)$  for the aforesaid age groups, further divided by sex} (Men, Women,  Both Sexes) \textbf{and health conditions} ( "Healthy", and "General" ).  In particular:
\begin{itemize}

\item \textbf{Figure 1} shows the \textbf{dependence on age of the BMP average values}, respectively for "Men Only" (left side histogram), "Both Sexes" (central histogram), "Women Only" (right side histogram), distinguishing within each picture between the "healthy" and the "general" cases.

\item	\textbf{Figure 2} shows, \textbf{for the "General" case}, the \textbf{dependence on age of the spreading of the BMP values around their average}, respectively for "Men Only" (left side), "Both Sexes" (center), "Women Only" (right side): \textbf{each  result is presented as a $2.\sigma$ vertical bar centered on the  average value AV}.

\item \textbf{Figure 3} shows the corresponding results of the figure above for the \textbf{HP case}.
\end{itemize}
\begin{figure}[h]\label{figura23}
\begin{center}
\includegraphics[width=5.5cm]{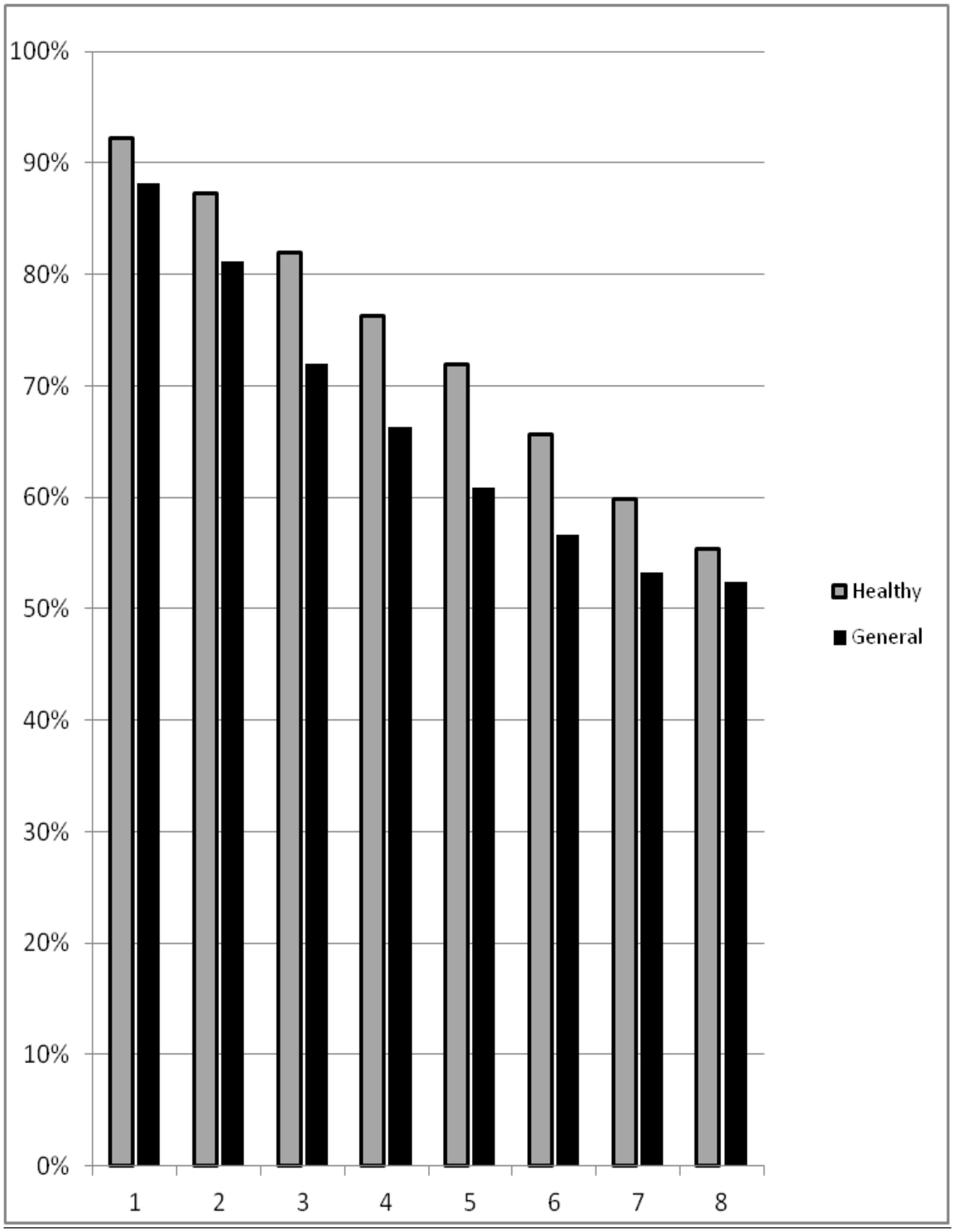}
\includegraphics[width=5.5cm]{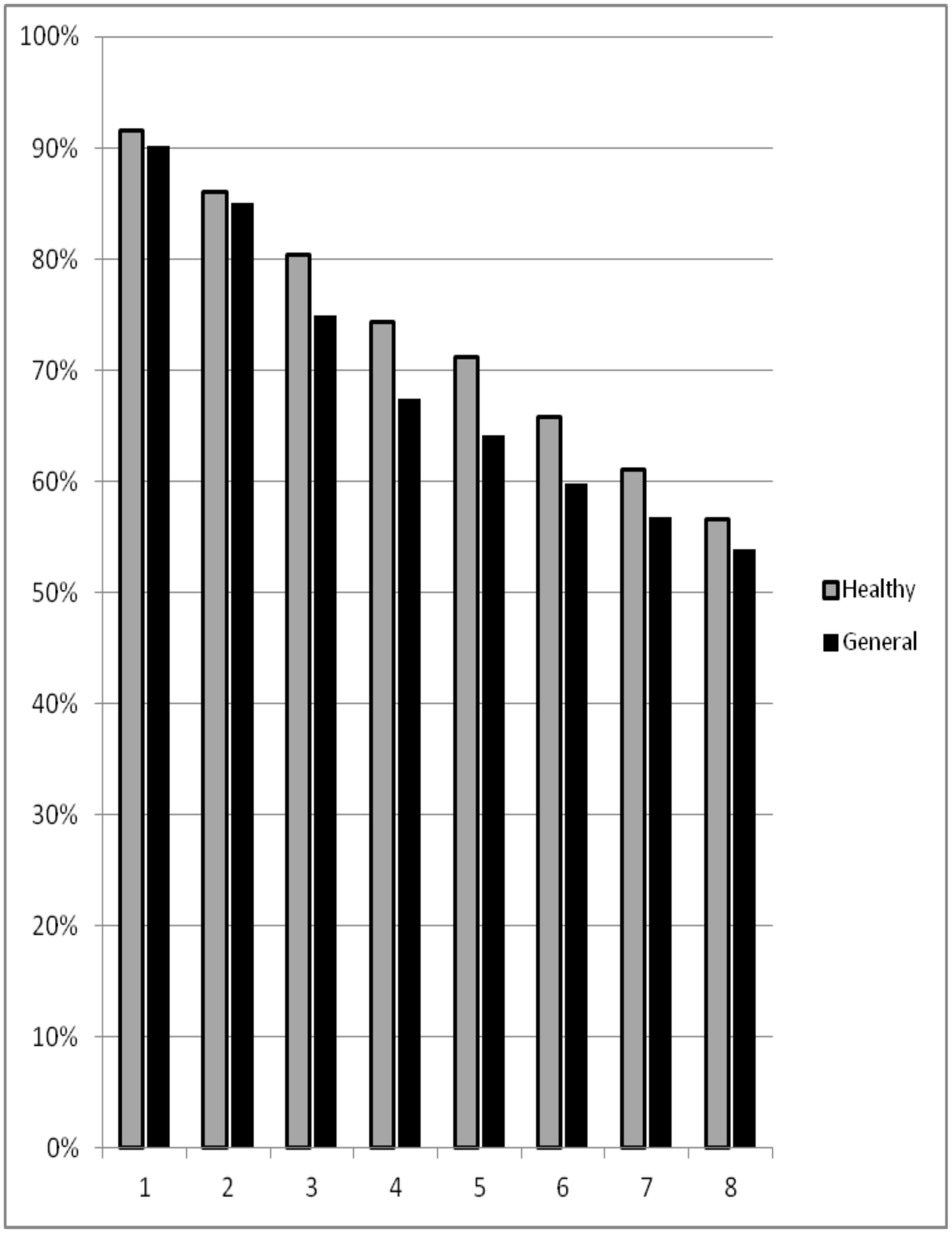}
\includegraphics[width=5.5cm]{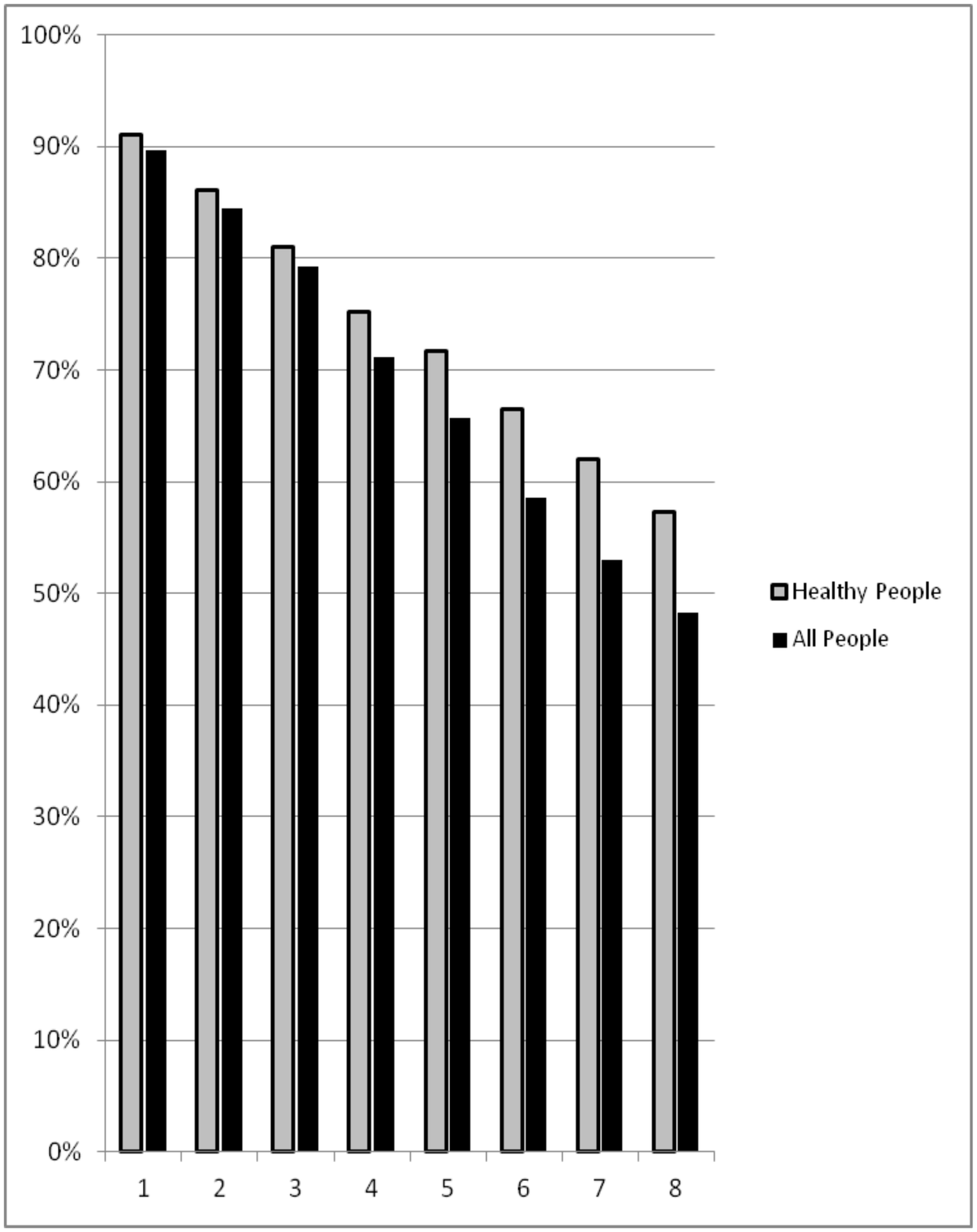}
\caption{BMP (percent) Average Values Vs Age Groups: MEN (Left), BOTH SEXES (center), WOMEN (right).}
\end{center}
\end{figure}
\begin{figure}[h]\label{figura56}
\begin{center}
\includegraphics[width=5.5cm]{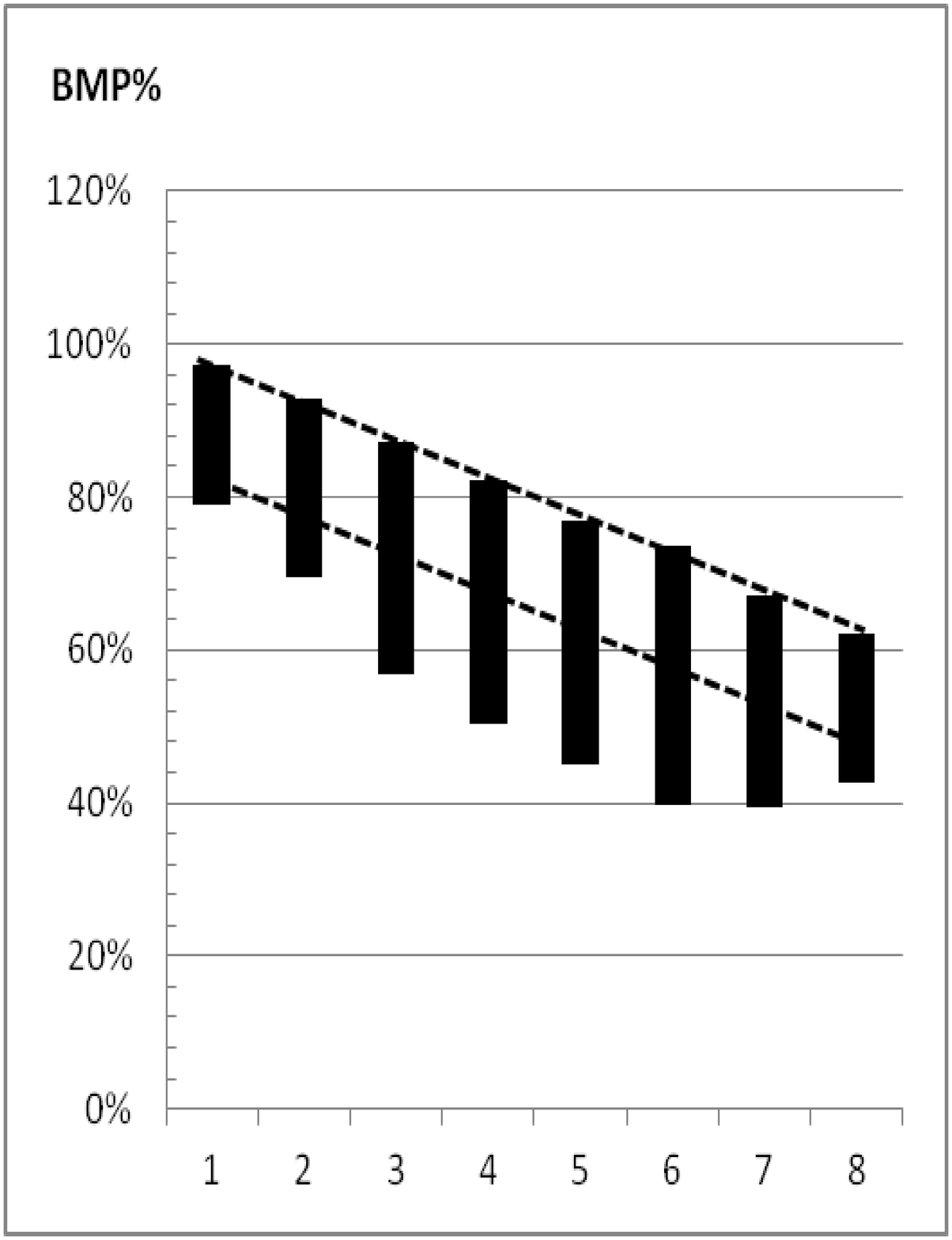}
\includegraphics[width=5.5cm]{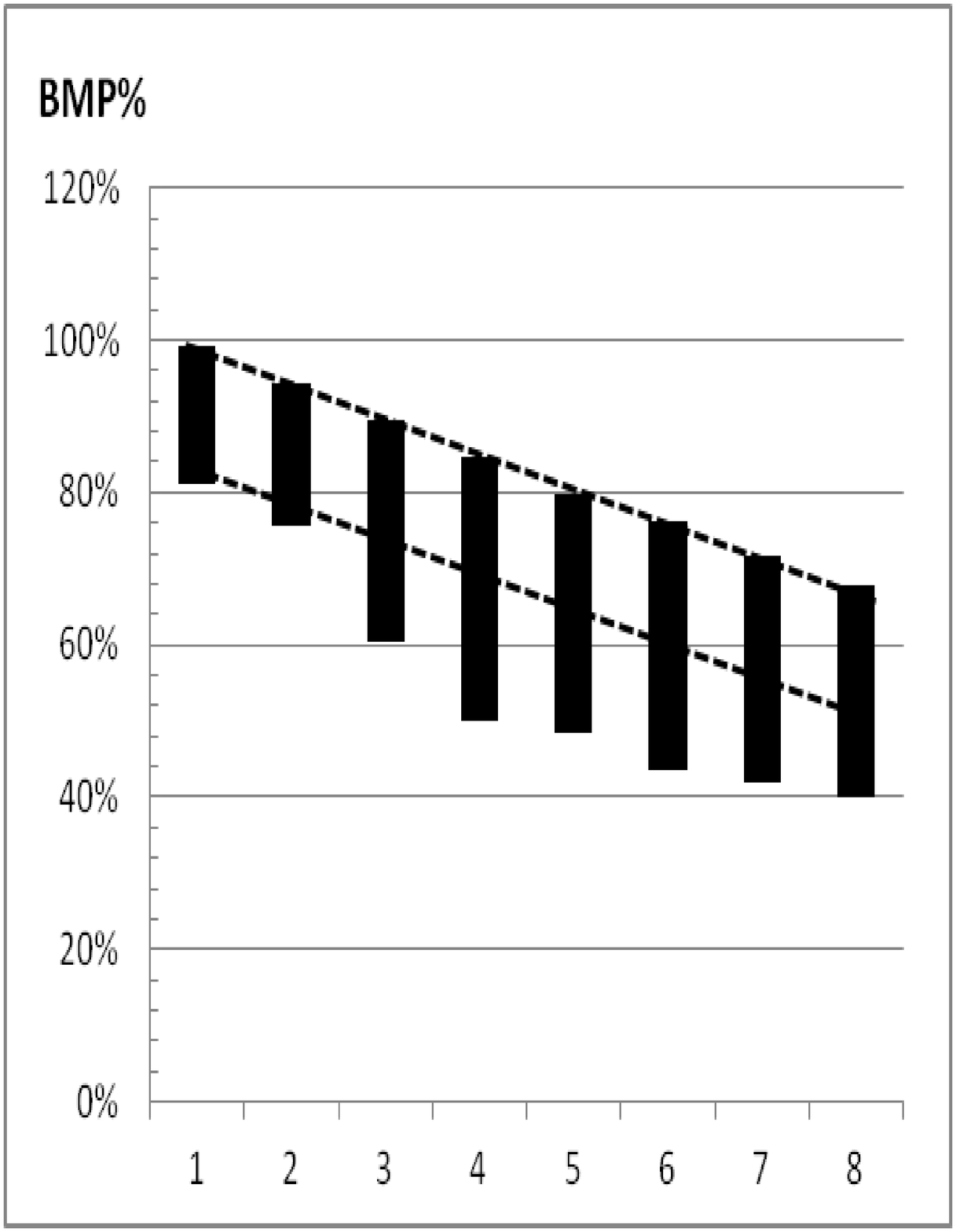}
\includegraphics[width=5.5cm]{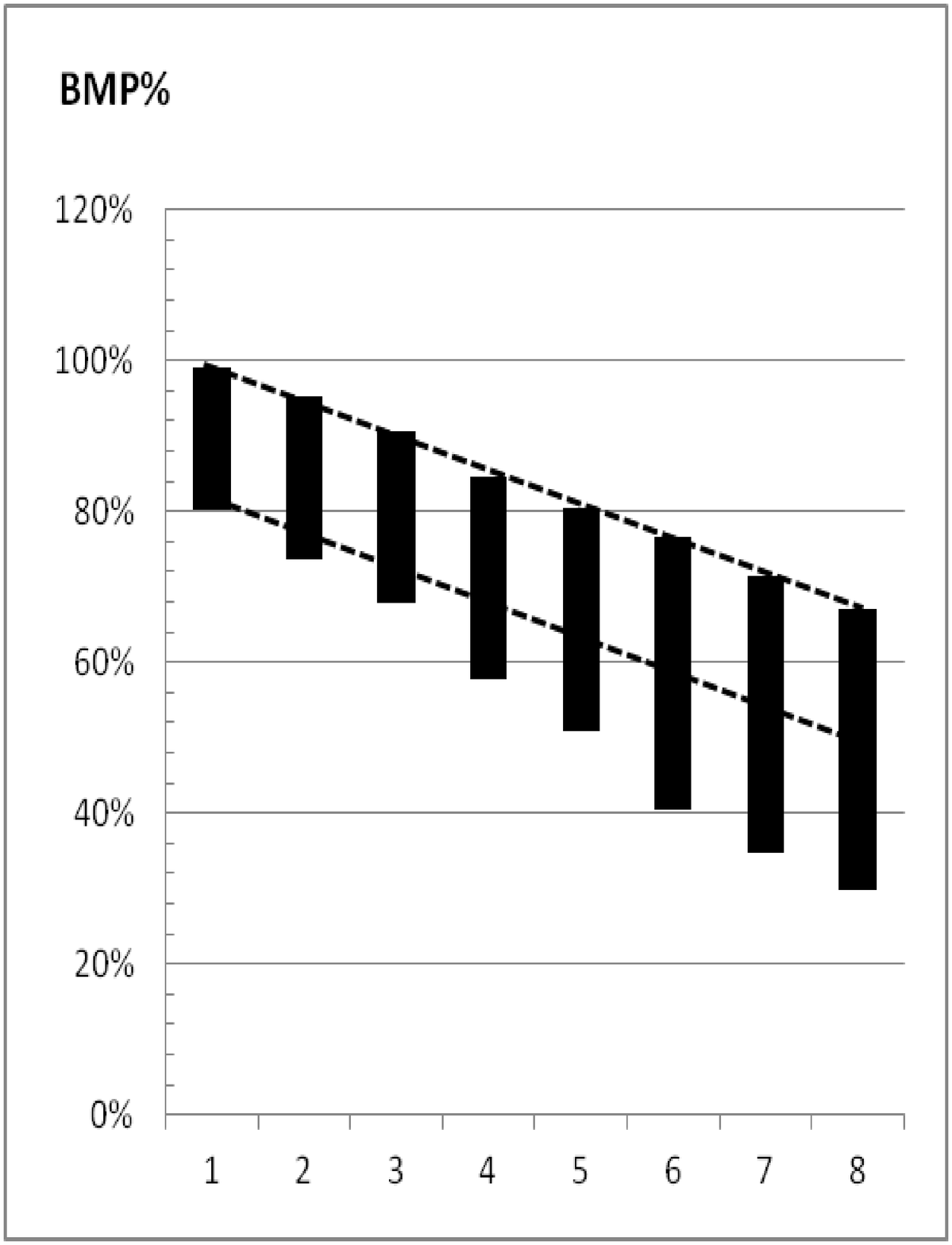}
\caption{BMP (percent) Vs Age Groups in the "GENERAL CASE": MEN (left), BOTH SEXES (center), WOMEN (right).}
\end{center}
\end{figure}
\begin{figure}[h]\label{figura89}
\begin{center}
\includegraphics[width=5.5cm]{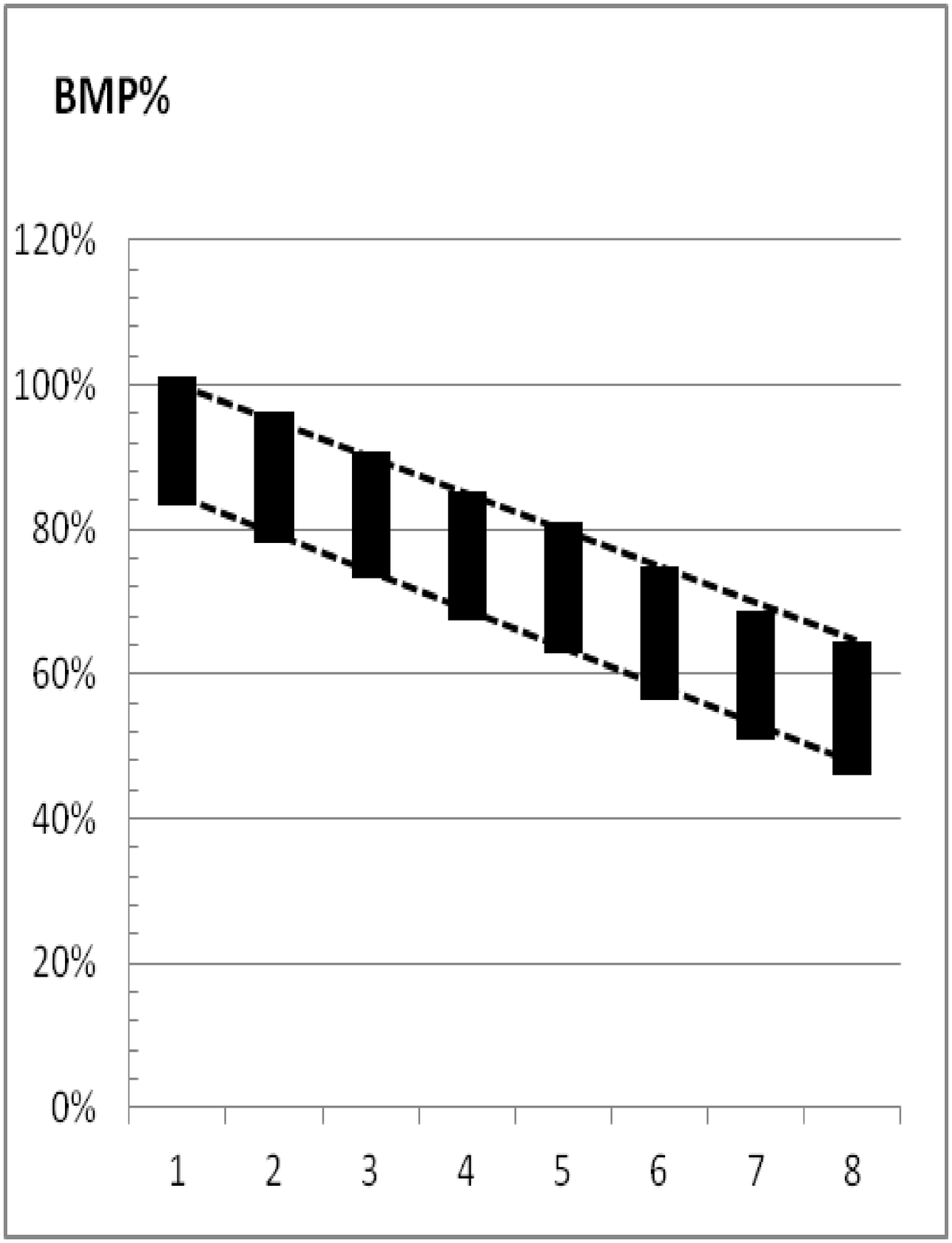}
\includegraphics[width=5.5cm]{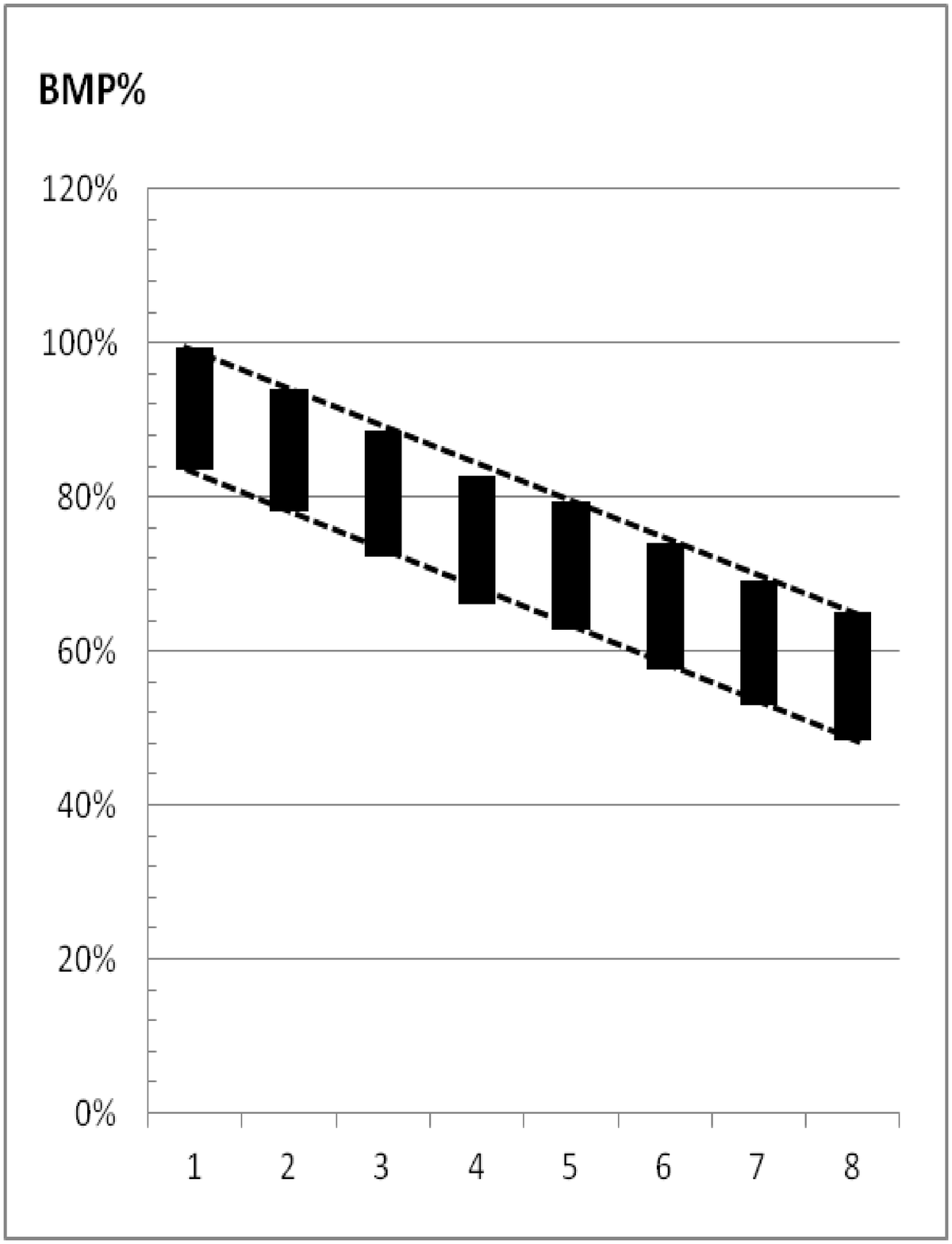}
\includegraphics[width=5.5cm]{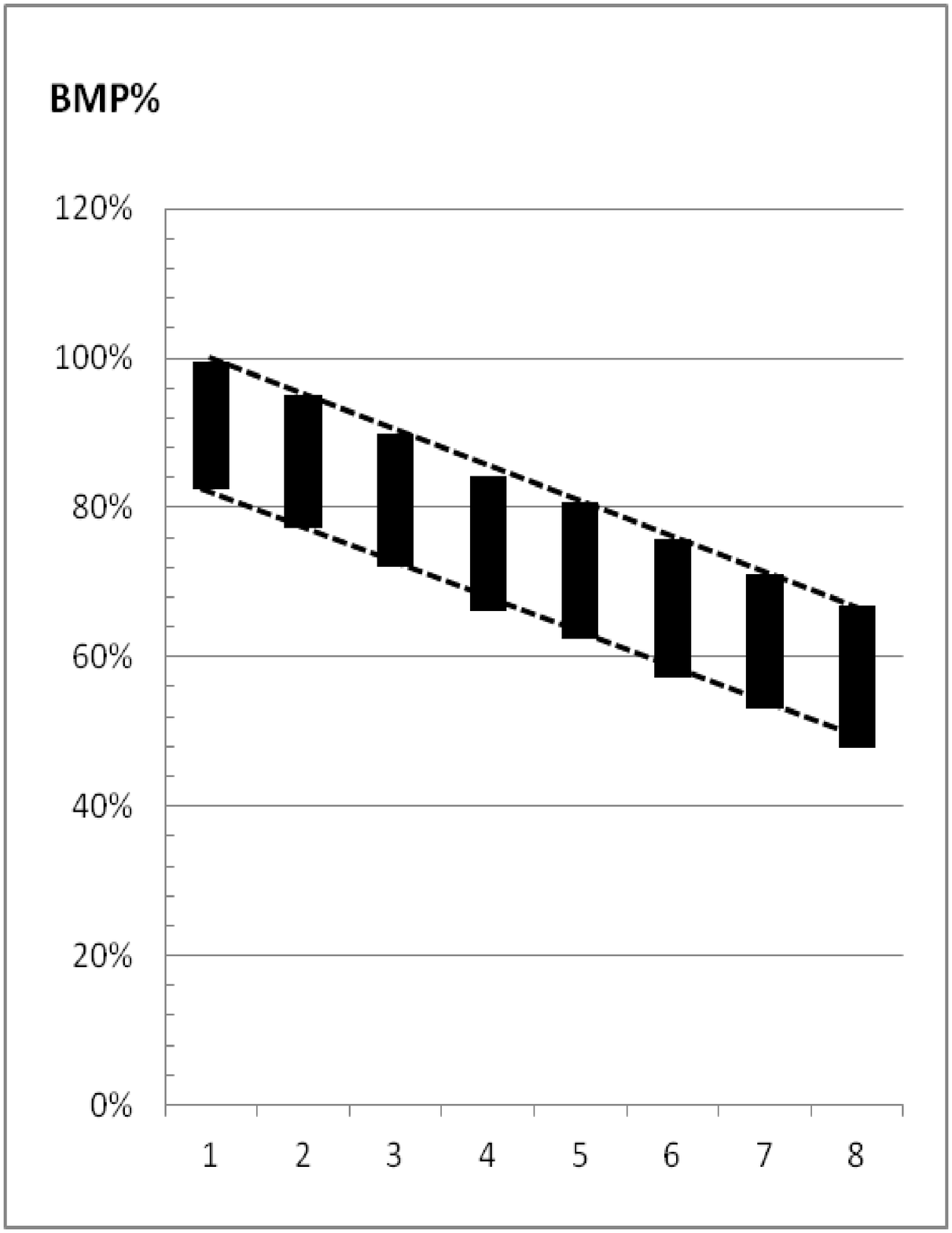}
\caption{BMP (percent) Vs Age Groups for Healthy People Only: "MEN" (left), "BOTH SEXES" (center), "WOMEN" (right).}
\end{center}
\end{figure}
By analyzing these results, the following  macroscopic - and partially unexpected - considerations are self-evident:
\begin{itemize}
\item \textbf{BMP values decrease as the age of the patient increases}, for all the classes considered  (both sexes, men, women), \textbf{thus a BMP decay is to be considered  normal and natural} for both the healthy and the non-healthy: this agrees with similar conclusions, available in the literature,  about  the dependence of the HRV circadian profile on age\cite{32}.

\item	Within each class , \textbf{the decrease of the BMP distribution is different according to whether only the healthy or the general  population is considered}.

\item	In particular, the BMP distributions of values decrease with age almost \textbf{linearly and with constant standard deviation values  for the HP groups}, whereas their decreases become steeper,  and with growing standard deviation values,  \textbf{in the other cases}.

\item	Finally, taking into account that the dashed lines in Figure 2 report for comparison the same linear decay of the corresponding BMP values represented for the HP in Figure 3, it is evident that the spreading below the two dashed lines refers to "non-healthy" people: \textbf{thus an illness status seems to induce a faster decay of BMP}.
\end{itemize}

According to the latter consideration, by evaluating the slopes of the linear decline from the data regarding HP in Figure 3,  it is easy to find that the decay is given by:
\begin{center}

\textbf{BOTH SEXES :	Average Yearly BMP Natural Decay $= - 0.500 \pm 0.021$ percent/year}

\bigskip

\textbf{MEN:		Average Yearly BMP Natural Decay $= - 0.478 \pm 0.010$ percent/year}

\bigskip

\textbf{WOMEN:	Average Yearly BMP Natural Decay $= - 0.528 \pm 0.047$ percent/year}
\end{center}
The numbers above lead to a sort of \textbf{general rule}, immediately useful even though simplified and approximate: an average \textbf{yearly BMP decay of  0.5percent/year  is natural and normal for healthy  people} ( 4-5 percent more for women, 4-5 percent less for men); \textbf{a larger yearly decay might indicate pathologies to be investigated}; a very large decay could indicate \textbf{severe pathologies}, or, in any case, a \textbf{process of biological aging much faster} than the average. Furthermore, \textbf{the accelerated decay process} induced by pathologies \textbf{seems to be different for male and female genders. Female BMPs} seem to be \textbf{more stable} for a longer period, presenting \textbf{a faster decay in the later stage of life}, whereas \textbf{male BMPs} show \textbf{a rapid decay in middle-age}, to remain stable until the later stage.

\begin{figure}\label{tabellaA}
\begin{center}
\includegraphics[width=6.5cm]{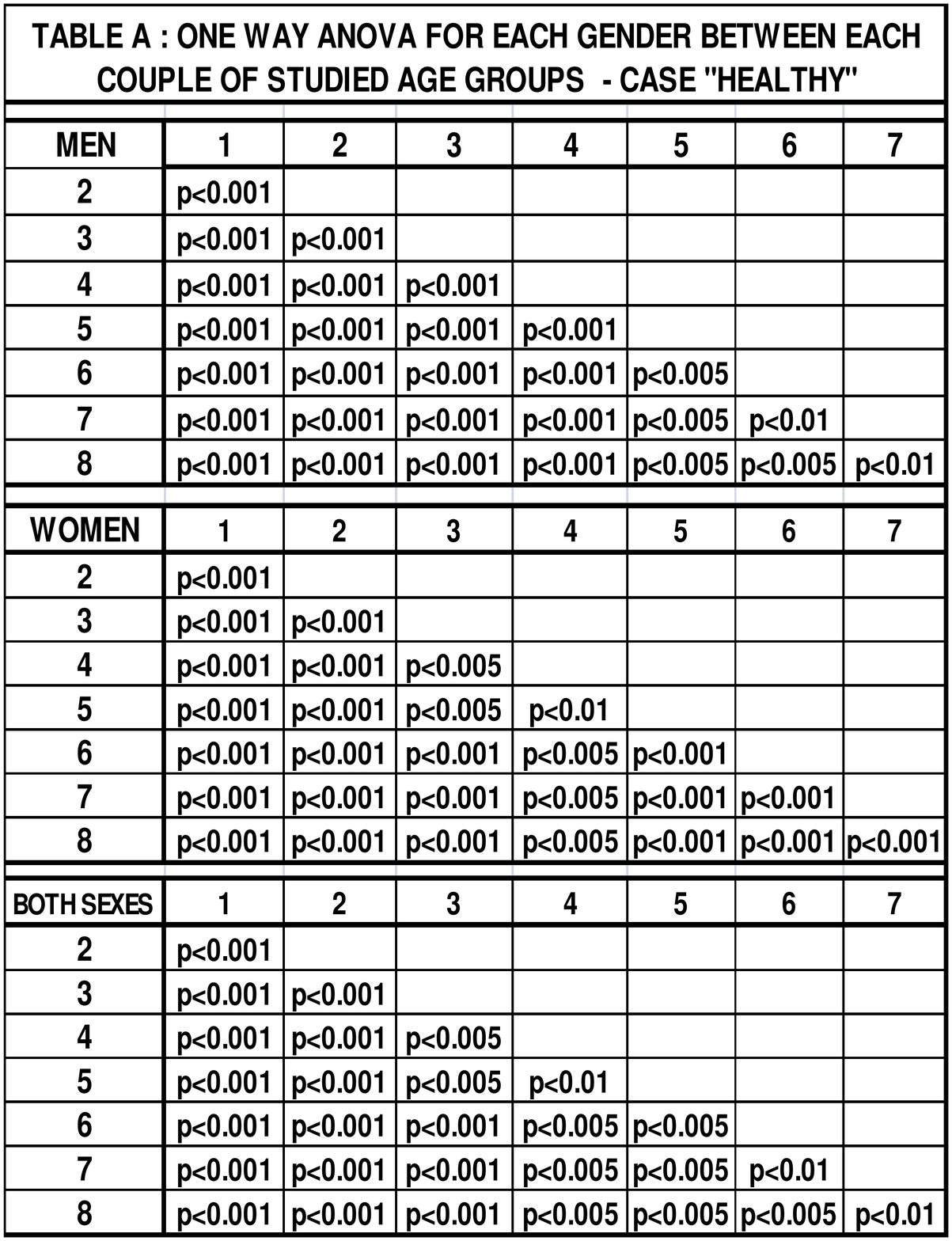}
\includegraphics[width=6.5cm]{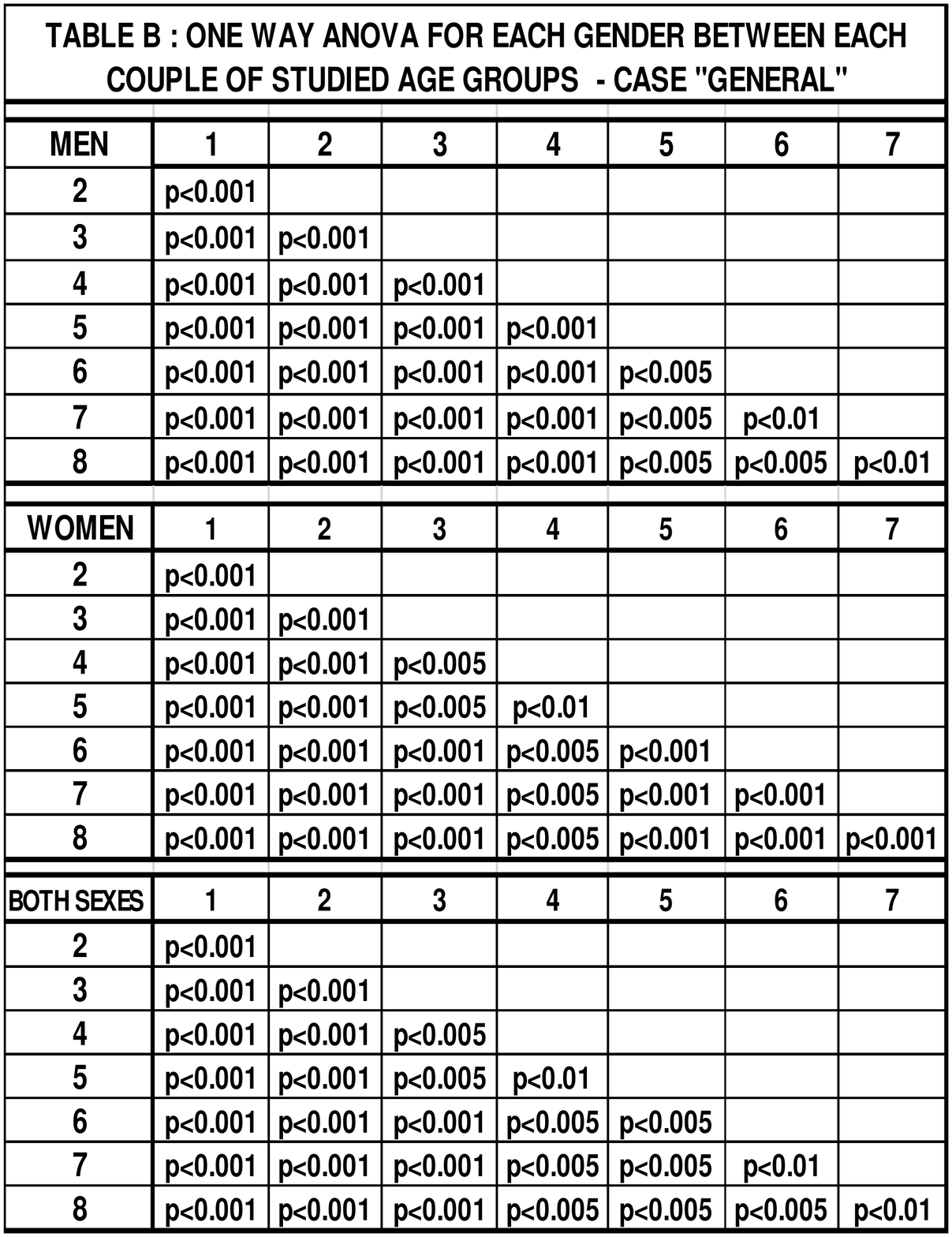}
\includegraphics[width=4.0cm,height=8.43cm]{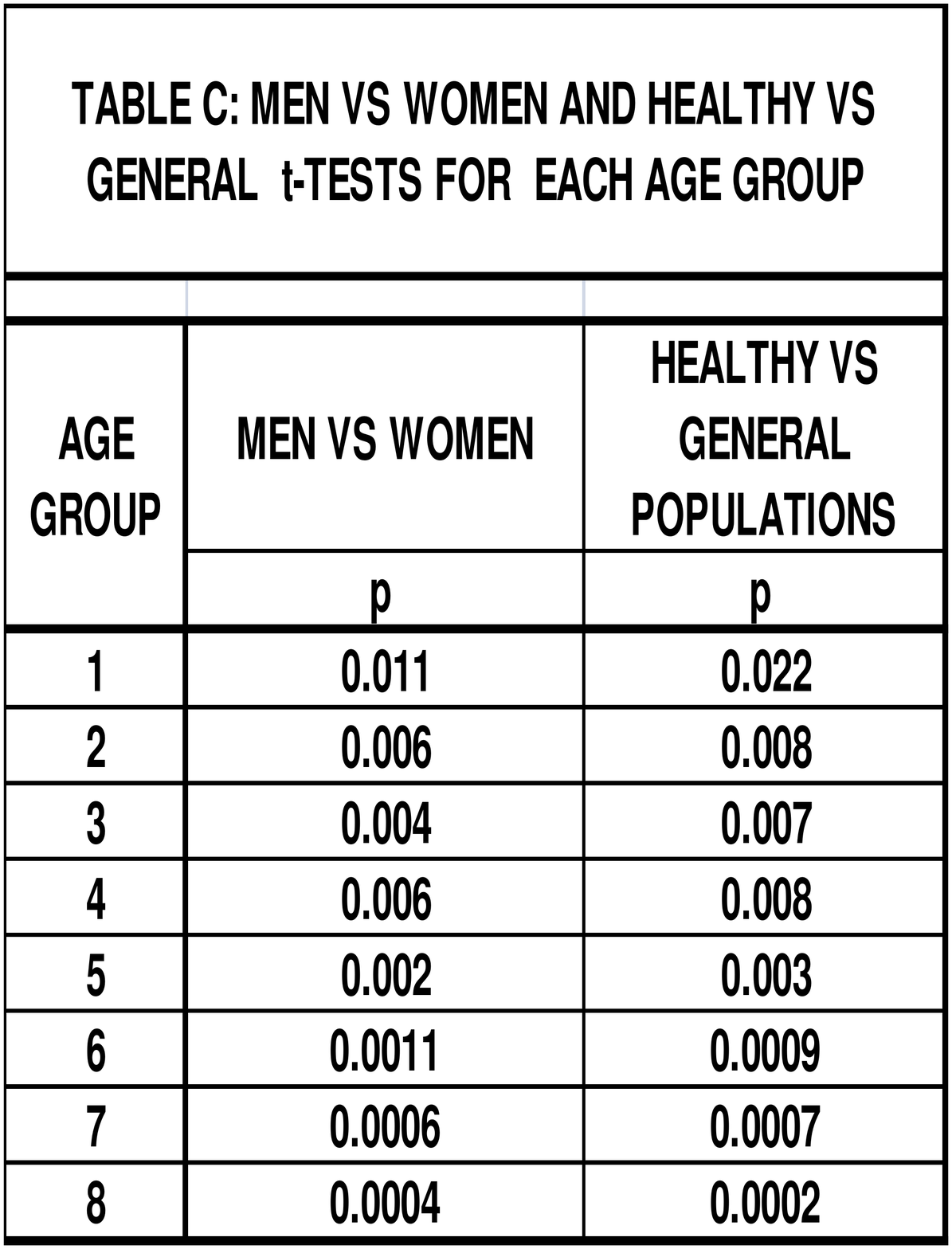}
\caption{Tables A,B,C respectively}
\end{center}
\end{figure}

While carrying out  this research work, the whole set of linear and non-linear variables - such as SDANN, RMS-SD, pNN50, TIN, VLF  LF and HF Powers, LF/HF Ratio, Poincare' SD1 and SD2, Shannon Entropy SE- was calculated  from the HRV data. It is important to remark that, unlike BMP, \textbf{none of these parameters shows any form of regular dependence on  the patient's age}, and, therefore, only BMP can be expected to provide \textbf{cognitive or predictive  information} from this kind of age and gender analyses. Nevertheless a further function  $\Theta$ (SDANN,..., LF/HF,...,SE) , can be calculated with the linear and non-linear variables above and added to  BMP, thus providing a set of "Extended BMP Values" which is very promising for diagnostic purposes. This extension is not dealt with in the present paper,  but it will be probably considered in a future paper.

\bigskip

\textbf{As far as statistical analysis is concerned}, Kolmogorov-Smirnov test was used to assess the normality of the distribution of all  variables. A one-way ANOVA, with Turkey's multiple comparison test, was performed to assess statistical differences in the BMP values between each couple of populations, separately for the healthy and the general cases and for each gender. Results are shown respectively in \textbf{TABLES A} (Healthy) and \textbf{B} (General)  and \textbf{significant differences in all cases are well demonstrated}. Furthermore, a paired t-test was performed to evaluate differences between genders and between the "healthy" and the "general" within each age group. Results, referred in \textbf{TABLE C}, show clear differences for the higher age groups, whereas minor differences appear for the younger age groups, as it was predictable because of the common "good health" conditions in those populations.
\section{Conclusions}

The results given in Sec. $III.A$  define the values of the reference parameter $BMP_H^{AV}$ for  the "standard case" , thus providing a way to determine the BMP dependence on  age and gender, as shown Sec. $III.B$.

\bigskip

The results shown in Sec. $III.B$  seem to be of the highest interest, and justify the name of "\textbf{life potential}" for the BMP algorithm: \textbf{it assumes, by definition, values close to 100 percent for  healthy people} (when the whole life potential is still unexploited) \textbf{and decreases when age increases and/or health conditions become critical} (so that the residual life potential vanishes step by step).

\bigskip

Therefore BMP is now well established for  cognitive analyses and it is likely that in the near future it could also be used for diagnostic/prognostic clinical purposes. This is the main task of the on-going phase of research  , which will  also take into account cross analyses of BMP with BRSA\cite{24,25}  (Baro-Reflex Sensitivity Analysis), PDSP\cite{26,30}  (Phonocardiography Digital Signal Processing) and RIAC\cite{31} (Radio-Isotopic Angio-Cardiography).

\bigskip

POLISA, University of Calabria and Ascoli Piceno Hospital will welcome any set of  24 hours raw RR data, supplied in any accessible format (ASCII, EXCEL, etc) and are interested in processing such data  to  evaluate the "life potential"  BMP and other linear and non-linear variables;  they are also willing to cooperate  with other scientific Institutions in order to  increase the size of the data base and therefore the reliability of the results.

\section*{Acknowledgements}

The Authors are grateful  to \textbf{Mortara Inc}. for the help provided   in  solving interface problems between MORTARA and POLISA software. They are also indebted to \textbf{Prof. E. Pugliese Carratelli} for  helpful discussion  on some  mathematical issues.



\begin{thebibliography}{9}

\bibitem{1}	Akselrod S, Gordon D, Ubel FA, Shannon DC, Berger AC, Cohen RJ. {\em Power spectrum analysis of heart rate fluctuation: A quantitative probe of beat-to-beat cardiovascular control}. Science. 1981; \textbf{213}:220-222.

\bibitem{2}	Heart rate variability. Standards of measurement, physiological interpretation, and clinical use. Task force of the European society of cardiology and the North American society of pacing and electrophysiology. Eur Heart J. 1996;\textbf{17}:354-381.

\bibitem{3}	Barra O.A., Moretti L., HRV Non-Linear Parameters and Crossed Analyses HRV-BRSA-PDSP-RIAC, Research Reports 2012, 10. Politecnico Internazionale "Scientia et Ars", Vibo Valentia , Italy.

\bibitem{4}	Steeb W.H., A Handbook of Terms used in Chaos and Quantum Chaos, BI-Wissenschaftsverlag, Mannheim, 1991.

\bibitem{5}	Schuster H.G., Deterministic Chaos, Physik-Verlag, Weinheim, 1984.

\bibitem{6}	Shannon C., Weaver W., The Mathematical Theory of Communication, University of Illinois Press, Urbana, 1948.

\bibitem{7}	Zaslavsy G.M., Chaos in Dynamic Systems, Harwood, Chur, 1985.

\bibitem{8}	Mandelbrot B.B., The Fractal Geometry of Nature, Freeman, San Francisco, 1982.

\bibitem{9}	Yamamoto Y, Hughson RL. Coarse-graining spectral analysis: new method for studying heart rate variability. J Appl Physiol. 1991;\textbf{71}:1143-1150.

\bibitem{10}	Babloyantz A, Destexhe A. Is the normal heart a periodic oscillator? Biol Cybern. 1988;\textbf{58}:203-211.

\bibitem{11} Saul JP, Albrecht P, Berger RD, Cohen RJ. Analysis of long-term heart rate variability: methods, 1/f    scaling and implications. In: Computers in Cardiology 1987. IEEE Computer Society Press .Washington, DC; 1988:419-422.

\bibitem{12} Brennan M,Palaniswami M, Kamen P. Do existing measures of Poincaré plot geometry reflect non-linear features of heart rate variability? Proc. IEEE Transactions on Biomedical Engineering, 2001, \textbf{48}, 1342-1347

\bibitem{13} Kanters JK, Holstein-Rathlou NH, Agner E (1994). "Lack of evidence for low-dimensional chaos in heart rate variability". Journal of Cardiovascular Electrophysiology \textbf{5} (7): 591-601.

\bibitem{14} Storella RJ, Wood HW, Mills KM et al. (1994). "Approximate entropy and point correlation dimension of heart rate variability in healthy subjects". Integrative Physiological and Behavioral Science \textbf{33} (4): 315-20.

\bibitem{15} Carrasco S., M.J. Cait´an, R. Gonz´alez, and O. Y´anez. Correlation among Poincar´e plot indexes and time and frequency domain measures of heart rate variability. J Med Eng Technol, 2001, \textbf{25}(6):240-248.

\bibitem{16} Fusheng Y., Bo H., and Qingyu T., Approximate entropy and its application in biosignal analysis. In M. Akay, editor, Non-linear Biomedical Signal Processing: Dynamic Analysis and Modeling, volume II, , IEEE Press, New York, 2001, pages 72-91.

\bibitem{17} Kobayashi M, Musha T. 1/f fluctuation  of  heart beat period.  IEEE Trans Biomed Eng. 1982;\textbf{29}:456- 457

\bibitem{18} Guzzetti S., Signorini M.G., Cogliati C., Mezzetti S., Porta A., Cerutti S., and Malliani A., Non-linear dynamics and chaotic indices in heart rate variability of normal subjects and heart-transplanted patients. Cardiovascular Research, 1996, \textbf{31}:441-446.

\bibitem{19} Henry B., Lovell N., and Camacho F., Non-linear dynamics time series analysis. In M. Akay, editor, Non-linear Biomedical Signal Processing: Dynamic Analysis and Modeling, volume II,. IEEE Press, New York, 2001; pages 1-39.

\bibitem{20}	Lake D.E., Richman J.S., Griffin M.P., and Moorman J.R., Sample entropy analysis of neonatal heart rate variability. AJP , 2002, \textbf{283}:R789-R797.

\bibitem{21}	Penzel T., Kantelhardt J.W., Grote L., Peter J-H., and Bunde A., Comparison of detrended fluctuation analysis and spectral analysis for heart rate variability in sleep and sleep apnea. IEEE Trans Biomed Eng, 2003, \textbf{50}(10):1143-1151.

\bibitem{22} Richman J.A. and Moorman J.R., Physiological time-series analysis using approximate entropy and sample entropy. Am J Physiol, 2000, \textbf{278}:H2039-H2049.

\bibitem{23}	Webber C.L. Jr. and Zbilut J.P., Dynamical assessment of physiological systems and states using recurrence plot strategies. J Appl Physiol, 1994, \textbf{76}:965-973.

\bibitem{24}	La Rovere MT, Bigger JT, Marcus FI, Mortara A, Camm AJ, Hohnloser SH, Nohara R, Schwartz PJ on behalf of the ATRAMI Investigators: Prognostic value of depressed baroreflex sensitivity.  The ATRAMI Study. Circulation 1995, \textbf{92} (supp. I),I-676.


\bibitem{25}	Bigger JT Jr, La Rovere MT, Steinmann RC, Fleiss JL, Rottman JN, Rolnitzky LM, Schwartz PJ. Comparison of baroreflex sensitivity and heart rate variability after myocardial infarction.
J Am Coll Cardiol 1989; \textbf{14}: 1511-1518

\bibitem{26}	R. L. Allen and D. W. Mills,  Signal analysis: time, frequency, scale and structure. New York, Piscataway, N.J.: Wiley; IEEE Press, 2004.

\bibitem{27}	R. N. Bracewell, The Fourier  transform and its applications, 3. ed. Boston: McGraw Hill, 2000.

\bibitem{28}	T. Olmez and Z. Dokur, "Classification of heart sounds using an artificial neural network", Pattern Recogn. Lett., 2003,  \textbf{24}, pp. 617-629.

\bibitem{29}	L. G. Durand and P. Pibarot, "Digital signal processing of the phonocardiogram: review of the most recent advancements", Crit. Rev. Biomed. Eng., 1995, \textbf{23}, pp.163-219.

\bibitem{30}	R. M. Rangayyan and  R. J. Lehner,  " Phonocardiogram  signal analysis: a review", Crit Rev Biomed Eng,  1987, \textbf{15}, pp. 211-236.

\bibitem{31}	Mohaved A., Gnanasegaran G., Buscombe J. And Hall M., "Integrating Cardiology for Nuclear MedicinePhysicians" , Springler-Verlag Berlin Heidelberg, 2009.

\bibitem{32}	Vandeput S., Verheyden B., Aubert A.E. and Van Huffel S.m "Non-Linear Heart Rate Dynamics: Circadian Profile and Influence of Age and Gender", Med. Eng.and Phys., 2012, \textbf{34}, pp. 108-117.


\end{thebibliography}


\end{document}